\documentclass[10pt,prd,aps,twocolumn,nofootinbib,
preprintnumbers,superscriptaddress]{revtex4-2}

\usepackage{natbib}
\usepackage{amssymb,amsbsy,amsmath,amsfonts}
\usepackage{mathrsfs}
\usepackage{color}
\usepackage{graphicx}
\usepackage{slashed}
\usepackage{braket}

\newcommand{\nn}{\nonumber}
\renewcommand{\Im}{{\rm Im\,}} 
\renewcommand{\Re}{{\rm Re\,}}

\newcommand{\be}{\begin{equation}}
\newcommand{\ee}{\end{equation}}

\usepackage{cmbright}		

\begin{document}

\title{Semi-inclusive pion electroproduction at the highest transverse momenta}

\author{Andrei Afanasev}
\affiliation{Department of Physics,
The George Washington University, Washington, DC 20052, USA}

\author{Carl E. Carlson}
\affiliation{Physics Department, William \& Mary, Williamsburg, Virginia 23187, USA}

\date{\today}

\begin{abstract}
   At the energies of present and future electron accelerators designed to study the structure of hadrons, there is a regime where hard pion electroproduction proceeds by a perturbatively calculable process in QCD. The process  is not the leading twist fragmentation one but rather a higher twist process that produces kinematically isolated pions. Semi-inclusive data may teach us more about parton distribution functions of the target and the pion distribution amplitude. In addition, there is a connection to generalized parton distribution calculations of exclusive electroproduction of mesons in that the perturbative kernel is the same.
\end{abstract}

\maketitle


\section{Introduction}


This article studies meson production at the highest transverse momentum under given kinematics of deep-inelastic lepton-proton (or other hadronic target) collisions.   The focus will be on direct pion production, and will include discussion of other production processes, will show numerical results, and will explore what can be learned from observations of directly produced pions.

Direct pions, also referred to as isolated pions, are pions produced within a short distance of the initiating reaction in processes that are perturbatively calculable. The name isolated follows because the production process creates only one particle, the pion (or meson, more generally), with significant momentum in its direction.  There will be a recoil quark traveling with significant momentum in an azimuthally opposite direction and some low momentum debris from target fracture, but the pion will be traveling alone and in particular will not be part of a jet.

Direct pion production was first suggested in~\cite{Baier:1980yk} and there have been a number of studies since, for examples see~\cite{Carlson:1991es,Carlson:1993ys,
Brandenburg:1994mm,Afanasev:1999xk,Afanasev:2003ne,Liu:2019srj}.  The present, however, is a good time for further study as it appears that data where direct meson production is the dominant process are within reach at Jefferson Lab and the Electron-Ion Collider under construction at BNL.  


In addition, when direct pion production is observed, the measured cross sections can teach us more about the structure of both the target and the produced meson.  Specifically, one can make choices of kinematic ranges where the pion structure does not impact the falloff of the cross section with increasing, for example, pion transverse momentum,  but that falloff is nonetheless directly proportional to the momentum fraction ($x$) dependence of the target's quark distribution amplitude (pdf).  Hence one gets a measure of that pdf.  The measurement will be at high $x$ where the current experimental data does not give the pdf with high precision (although lattice calculations are helpful here~\cite{HadStruc:2021qdf,JeffersonLabAngularMomentumJAM:2022aix,HadStruc:2022nay}).  Other choices of kinematics conversely can give shape insensitivity to the quark pdf and allow measuring some particular momentum transfer dependent integral over the pion's distribution amplitude (DA), which with enough data can be inverted to give the distribution amplitude itself (a feature already noted in~\cite{Baier:1980yk}).

We will make some general remarks on kinematics in the next section, and show the calculation of the isolated pions in some detail in the following section.  In Sec~\ref{sec:other} we give, for the purpose of comparison, an outline of the fragmentation and vector meson dominance processes, attempting to be brief but with enough detail to allow numerical calculation.  Then in Sec.~\ref{sec:numbers}, we give a selection of numerical results. We start with a comparison of the sizes of the different processes at various transverse momenta and various photon virtuality, demonstrating that there indeed exists a region where direct pion production should be dominant. Then for the direct pion process itself, we show results for the different  structure functions that appear in the cross section expression for pion electroproduction, as in Eq.~\eqref{eq:xsctn} below.  Section~\ref{sec:learn} shows examples of ways to  extract information on parton distribution functions (pdf) and pion distribution amplitudes from direct pion data.  Some closing remarks are offered in Sec.~\ref{sec:end}.


\section{General Kinematic Review}
\label{sec:kin}

\begin{figure}[t!]
\includegraphics[width=0.7\columnwidth]{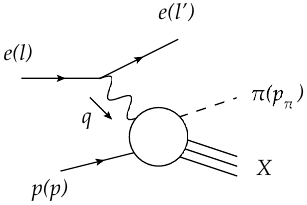}

\caption{Semiinclusive pion production from lepton proton scattering.} 
\label{fig:sidispi}
\end{figure}

Semiinclusive pion production from lepton-proton scattering, or semiinclusive deep inelastic scattering (SIDIS) where the observed hadron is a pion, is shown in Fig.~\ref{fig:sidispi}, with the lepton labeled "$e$" for definiteness.  The lepton part of the diagram can be calculated explicitly, and the full sixfold differential cross section is given in terms of 18 structure functions, in general.  For unpolarized protons only five structure functions enter, and~\cite{[][. One may also change the differential measures using $ d^3 \ell'/E' = 
m_p \nu \, dx \, dy \, d\psi $ and $d^3 p_\pi/\omega_\pi
= (\nu /2\omega_\pi ) \,
dz \, d\phi_h \, dp_{\pi\perp}^2$ (notation given in the reference).]Bacchetta:2006tn}
\begin{align}  \label{eq:xsctn}
E' E_\pi \frac{d^6 \sigma}{ d^3l' d^3p_\pi} 
&= \frac{\alpha^2}{ Q^4 } 
\frac{ 2 \omega_\pi }{ E (1-\epsilon) }
\left( 1 + \frac{m_p}{\nu} \right)
		\nn\\
&\hskip -2 em	\times \ 
	\Big\{F_{UU,T} + \epsilon F_{UU,L}
		+ \epsilon \cos(2\phi_h) F_{UU}^{\cos 2\phi_h}
            \nn\\
&	            
	+ \sqrt{ 2 \epsilon ( 1+\epsilon ) } \cos\phi_h \, F_{UU}^{\cos \phi_h}
            \nn\\
&
+ (2\lambda_e) \sqrt{ 2 \epsilon ( 1-\epsilon ) } \sin\phi_h \, F_{LU}^{\sin \phi_h}	 \Big\} .  
\end{align}
Here $E$, $E'$, and $E_\pi$ are the energies of the incoming and outgoing electron and of the pion, and $Q^2 = -q^2 > 0$ gives the four momentum of the virtual photon. 
Further, $\phi_h$ is the azimuthal angle between the lepton-photon and pion-photon scattering planes, $\lambda_e = \pm 1/2$ is the helicity of the incoming electron, and $\epsilon$ is the ratio of the longitudinal and transverse polarization probabilities of the virtual photon,
\begin{align}
\frac{1}{\epsilon} = \left( 1 +
  2\tau(1+\tau) \tan^2 (\theta_e/2) \right)
\end{align}
for lepton lab scattering angle $\theta_e$ and $\tau = \nu^2/Q^2$ where $\nu$ is the photon lab energy.  

The structure functions can be related to conventionally defined cross sections for $\gamma^* p \to \pi + X$,
\begin{align}
F_{mn} = \frac{ Q^2 (W^2-m_p^2)}{ 8 \pi^2 m_p \alpha}   \left(1+\frac{m_p}{\nu} \right)^{-1} 
\frac{d\sigma_{mn} }{ d^3 p_\pi }  ,
\end{align}
where $m,n$ are indices for the polarization vectors of the virtual photon and $W$ is the center-of-mass (CM) energy of the $\gamma^* p$.  There are also helicity indices for the proton target, which are here averaged over.   The photon longitudinal polarization is along $\vec q$\,; transverse polarization 
$\epsilon_1$ is in the photon-pion plane and in the direction of the pion's transverse momentum, and the other perpendicular to this plane, as $\epsilon_2 = \hat q \times \epsilon_1$.   Then
\begin{align}
&F_{UU,T} = \frac{1}{2} \left( F_{11} + F_{22}  \right) ,    \quad
F_{UU}^{\cos 2\phi_h} = \frac{1}{2} \left( F_{11} - F_{22}  \right) , \nn\\
&
F_{UU,L} = F_{LL} ,
\quad F_{UU}^{\cos \phi_h} = \Re F_{L1} ,	\quad	F_{LU}^{\sin \phi_h} = \Im F_{L1}	.
\end{align}
(For the virtual photon cross section we will make the conventional Hand flux factor choice~\cite{Hand:1963bb,Amaldi:1979vh,Arens:1996xw}.  The structure functions are independent of the choice, as witnessed by the factor $(W^2 - m_p^2)$ above.)

\begin{figure}[t!]
\includegraphics[width=0.8\columnwidth]{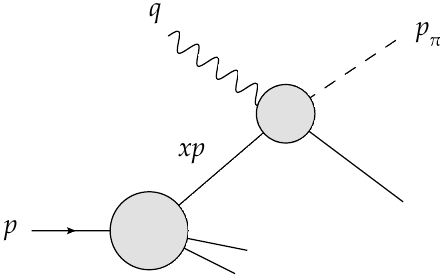}

\caption{Quark level diagrams for isolated pion production.} 
\label{fig:directbasic}
\end{figure}

\begin{figure*}[t!]
\includegraphics[width=1.0\textwidth]{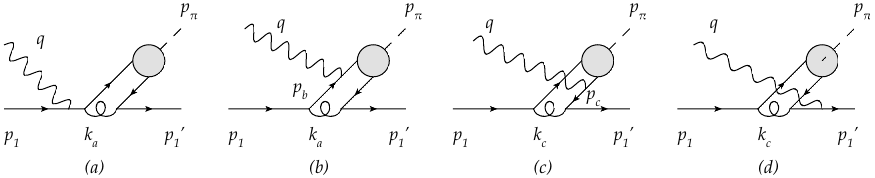}

\caption{Quark level diagrams for isolated pion production.} 
\label{fig:isodiagrams}
\end{figure*}



\section{Direct pion production}


Pions electro produced near the highest possible transverse momentum will be direct or  isolated pions. Isolated means that they travel in a given direction with no nearby hadrons of significant momentum; in particular, they are not part of a jet.  Fig.~\ref{fig:directbasic} illustrates the process.  The upper blob in this case can be perturbatively calculated from the diagrams in Fig.~\ref{fig:isodiagrams}.  The isolated pions come from processes where in addition to the photon being absorbed by a quark, there is also a hard gluon that produces a quark-antiquark pair, with the antiquark (most often) combining with the preexisting quark to produce a pion with the newly produced quark recoiling opposite to the pion's momentum, eventually combining with other matter from the target breakup to produce hadrons with no significant momentum in the direction of the fast pion.  


\subsection{Calculation}


The pions are formed from the nearly parallel quark and antiquark with momenta $y_1 p_\pi$ and $y_2 p_\pi = (1-y_1) p_\pi$.  They are connected by a Dirac factor $\gamma_5/\sqrt{2}$ and a pion distribution amplitude $\phi_\pi(y)$ and  there is an integral over $dy_1$.  Momentarily omitting the distribution amplitude and the integral, the first pair and second pair of diagrams give, respectively,
\begin{align}
\epsilon \cdot T_a &= -\frac{C_f e g^2}{\sqrt{2}}
\frac{2(2\lambda) e_1}{ k_a^2 p_b^2 }
\bar u(p'_1) \tau_a u(p_1)      \nn\\
&=   -\frac{C_f e g^2}{\sqrt{2}}
\frac{2(2\lambda) e_1}{ \hat s  t }
\frac{  \bar u(p'_1) \tau_a u(p_1)  }
    { y_2 ( y_1+y_2  q^2/t ) }  , \nn\\[1.2ex]
\epsilon \cdot T_c &= \frac{C_f e g^2}{\sqrt{2}}
\frac{2(2\lambda) e_2}{ \hat u  t }
\frac{  \bar u(p'_1) \tau_c u(p_1)  }
    { y_1 ( y_2+y_1  q^2/t ) }  .
\end{align}
The color factor is $C_f=4/(3\sqrt{3})$, the momenta can be read from the diagrams, $\lambda= \pm 1/2$ is the helicity of the incoming quark, $e_1$ and $e_2$ are quark charges in units of the elementary charge $e$, and the subprocess Mandelstam variables are
\begin{align}
\hat s &= (p_1 + q)^2 \,,   \nn\\
t &= (q - p_\pi)^2 \,,      \nn\\
\hat u &= (p_1 - p_\pi)^2 \,.  \nn\\
\end{align}
We treat the quarks and pions as massless, so that
\begin{align}
\hat s + t +\hat u = q^2  \,.
\end{align}

The $\epsilon$ are the virtual photon polarization vectors.  For transverse polarization $\epsilon_T$ with $T = 1,2$ we take $\epsilon_1$ in the subprocess scattering plane with the positive direction set by the transverse part of the pion momentum, and $\epsilon_2$ normal to the scattering plane.  

The Dirac matrices in context combine to
\begin{align}
\tau_a(\epsilon_T) &= \slashed{\epsilon}_T
\left( -y_1 \hat s + y_2 \hat u + q^2 \right)
    + 2 \slashed q \, y_2 \epsilon_T \cdot p'_1
    \,,     \nn\\
\tau_c(\epsilon_T) &= \slashed{\epsilon}_T
\left( + y_1 \hat s - y_2 \hat u + q^2 \right)
    - 2 \slashed q \, y_2 \epsilon_T \cdot p'_1
    \,.
\end{align}
We have used $\epsilon_T \cdot q = 0$, and $\epsilon_T \cdot p_1 = 0$, the latter valid in both the CM frame and the lab frame.  

The longitudinal polarization vector is
\begin{align}
\epsilon_L = \frac{1}{Q} \left( q
    - \frac{2q^2}{ \hat s - q^2 } p_1 \right),
\end{align}
where $Q=\sqrt{Q^2} = \sqrt{-q^2}$.  This polarization vector satisfies $e_L^2 = 1$ and 
$\epsilon_L\cdot q=0$.  Then
\begin{align}
\tau_a(\epsilon_L) &= 
- \frac{ 2 \slashed q q^2}{Q ( \hat s - q^2)}
    \,  y_2 \hat u   \,,   \nn\\
\tau_c(\epsilon_L) &= 
- \frac{ 2 \slashed q q^2}{Q ( \hat s - q^2)}
    \left(  y_1 \hat u + t \right)  .
\end{align}

To proceed to the ampitudes
\begin{align}
\epsilon \cdot \mathcal M
= \int_0^1 dy_1 \phi_\pi(y) \,
\epsilon\cdot\left( T_a + T_b \right) ,
\end{align}
we need integrals
\begin{align}
I_\pi J_1(q^2/t) &= \int_0^1 dy_1 
    \frac{ \phi_\pi(y) }{y_1+y_2 q^2/t} ,       \nn\\
I_\pi J_2(q^2/t) &= \int_0^1 dy_1 
\frac{ y_1 \phi_\pi(y) }{ y_2(y_1+y_2 q^2/t)} , \nn\\
J_2(q^2/t) &= 1 - \left(q^2/t \right) \, J_1(q^2/t).
\end{align}
The basic integral, used here as a normalization, is
\begin{align}
I_\pi = \int_0^1 dy_1 \frac{\phi_\pi(y)}{y_1}
\end{align}
and is the same integral that appears, for example, in the high $Q^2$ expression for the pion electromagnetic form factor
\begin{align}
F_\pi(Q^2) = 
    \frac{64 \pi \alpha_S}{3 Q^2} I_\pi^2  .
\end{align}

One finds
\begin{align}
\epsilon_T\cdot\mathcal M &= 
\frac{ \sqrt{\gamma} (2\lambda)}{ 2t^2 }
\bar u(p'_1) \left( A \slashed 
    \epsilon_T + B \, \epsilon_T \cdot p'_1 \,\slashed q \right) u(p_1) ,
            \nn\\
\epsilon_L\cdot\mathcal M &= 
\frac{ \sqrt{\gamma} (2\lambda)}{ - t }
\frac{ Q }{ \hat s - q^2 }
\bar u(p'_1) \, C \slashed q \, u(p_1) .
\end{align}
where
\begin{align}  \label{eq:crucialfactors}
A &= \frac{e_1}{\hat s} 
\left[ -t(\hat s -q^2) + J_1 \hat u (t-q^2) \right] \nn\\
& \hskip 0.5 em  + \frac{e_2}{\hat u}
\left[ t(\hat u -q^2) - J_1 \hat s (t-q^2) \right], \nn\\
B &= 2 t \left( 
\frac{e_1}{\hat s} J_1 + \frac{e_2}{\hat u} J_2 
        \right) ,        \nn\\
C &= J_1\left( 
\frac{e_1 \hat u}{\hat s} + \frac{e_2 \hat s}{\hat u}   \right)     
  + \frac{e_2 t}{\hat u} \left(J_1-1 \right)  .
\end{align}
The argument of $J_{1,2 }$ is $q^2/t$.

Also,
\begin{align}
\gamma = \frac{2^{13} }{27} \pi^3 I_\pi^2
    \alpha \alpha_S^2 
\end{align}
and it is useful to know
\begin{align}
    \epsilon_1 \cdot p'_1 =
\frac{\sqrt{\hat s t \hat u}}{\hat s - q^2}.
\end{align}

The subprocess cross section is 
\begin{align}
\frac{d\hat\sigma_{IJ}}{dt} = \frac{1}{16\pi(\hat s - q^2)^2}  
{\sum_{\text{q pol}}}'
(\epsilon_I\cdot\mathcal M_I)^* \ 
( \epsilon_J\cdot\mathcal M_J ) \,.
\end{align}
The prime indicates an average over the initial quark polarization, and $I,J = 1,2,L$ are chosen as appropriate to the cross section being calculated.  The various $d\hat\sigma$ may be understood to have suppressed indices stating the initial quark and the meson produced.


\subsection{Cross Section Formula Summary}


The quark level subprocess is embedded in a proton or other hadron, as in Fig. X for pion production from a virtual photon.  The Mandelstam variables for the virtual photon-proton process are measurable quantities and are
\begin{align}
s &= (p+q)^2  ,      \nn\\
t &= (q-p_\pi)^2  ,         \nn\\
u &= (p-p_\pi)^2  ,
\end{align}
and the momentum fraction of the struck quark can be obtained from these measurable quantities as
\begin{align}   \label{eq:xreal}
x = \frac{ -t }
{ s + u  - 2 m_p^2 - q^2}  \,.
\end{align}

The virtual photon cross section on the proton target is generically
\begin{align}
\frac{ d\sigma}{dx dt} = q_a(x)
    \frac{d\hat\sigma}{dt}(\gamma^* q_a \to \pi q_b)  ,
\end{align}
where an appropriate sum over quark flavors is implied;  $q_a(x)$ is the quark distribution function, or the probability of finding a quark of flavor $a$ with momentum fraction between $x$ and $x+dx$.  One can convert to other differentials using generically
\begin{align}
E_\pi \frac{d\sigma}{d^3p_\pi} = 
    \frac{2 m_p |\vec q\,| x^2}{-\pi t }
    \frac{ d\sigma}{dx dt}      \,,
\end{align}
where $| \vec q \,|$ is the virtual photon momentum in the lab frame. 

One can now give the structure functions, with prefactor
\begin{align}
 P_f = \frac{16\alpha_s^2 I_\pi^2}{27 \pi}
     \frac{ Q^2 (W^2-m_p^2) |\vec q\,| x^2}
         {t^4 \hat s (\hat s -q^2)^3} q_a(x)
\end{align}
are
\begin{align}
F_{UU,T} &= P_f
\Big\{      2A^2  (\hat s -q^2)^2   \nn\\
& \hskip 2 em - 2 AB \hat s \hat u (\hat s- q^2) + B^2 \hat s^2  \hat u^2      \Big\}   ,
                \\
F_{UU}^{\cos(2\phi)} &= P_f
\Big\{ - 2 AB \hat s \hat u (\hat s- q^2) + B^2 \hat s^2  \hat u^2      \Big\}  ,
                \\
F_{UU,L} &= - P_f
\ 8 q^2 \hat s t \hat u \, C^2     ,   
                \\
F_{UU}^{\cos\phi} &= P_f
\ 4 Q \sqrt{\hat s t \hat u} \, 
\left( A (\hat s-q^2) - B \hat s \hat u \right) C           ,
                \\
F_{LU}^{\sin\phi} &= 0.
\end{align}
The last follows because the lowest order perturbative amplitudes are all relatively real. It can of course become nonzero in higher order either through final-state gluon exchange or by combination with a T-odd PDF \cite{bacchetta2004new}.
Quantities $A$, $B$, and $C$ are given in Eqs.\eqref{eq:crucialfactors} in the previous subsection.

For comparison to photoproduction results, the 
$q^2 \to 0$ results are
\begin{align}
E_\pi \frac{d\sigma_T}{d^3p_\pi} &=
\frac{128\pi\alpha\alpha_S^2 I_\pi^2}{27} \,
\frac{2 m_p \omega_\gamma x^2}{t^2} q_a(x)
        \nn\\
&\hskip 1 em    \times 
\left( \frac{e_1}{\hat s} + \frac{e_2}{\hat u} \right)^2
\left( \hat s^2 + \hat u^2 \right)      ,
\end{align}
which agrees with with Eq.~(10) of~\cite{Afanasev:1997ie}.
Quantity $d\sigma_{TT}$ the same except for
$( \hat s^2 + \hat u^2 ) \to 2 \hat s \hat u$.


\subsection{Distribution amplitude integrals}


The distribution amplitude is normalized from the matrix element of the axial current between the pion and the vacuum.  The normalization is
\begin{align}
\int_0^1 dy_1 \, \phi_\pi(y) = \frac{1}{2\sqrt{3}} f_\pi
\end{align}
in the convention where $f_\pi \approx 93$ MeV.

Commonly used forms for the distribution amplitude are the asymptotic (or super-asymptotic) form and the Chernyak-Zhitnitsky form.  The asymptotic form is
\begin{align}
\phi_\pi(y)^{asy} = \sqrt{3} \, f_\pi y_1 y_2    \,,
\end{align}
from which
\begin{align}
I_\pi^{asy} = \frac{\sqrt{3}}{2} f_\pi
\end{align}
and
\begin{align}
J_1(b)^{asy} = \frac{ 1 -b^2 + 2b \log{b} }{(1-b)^3}.
\end{align}

An example of a broader distribution amplitude that gives larger $I_\pi$ than the asymptotic one,  is
\begin{align}     \label{eq:broad}
\phi_\pi(y)^{SR} = \frac{4}{ \pi \sqrt{3}} f_\pi
    \sqrt{ y(1-y) } \,,
\end{align}
whence
\begin{align}
I_\pi^{SR} &= \frac{2}{ \sqrt{3} } f_\pi \,,    \nn\\
J_1(b)^{SR} &= \frac{1}{ ( 1 + \sqrt{b}\, )^2 }  \,,
\end{align}
where SR stands for square root.

There is also the Chernyak-Zhitnitsky distribution amplitude which has a double hump structure and is given by
\begin{align}
\phi_\pi(y)^{CZ} = 5 \sqrt{3} \, f_\pi y_1 y_2 
(y_1-y_2)^2   \,.
\end{align}
Then
\begin{align}
I_\pi^{CZ} = \frac{5\sqrt{3}}{6} f_\pi
\end{align}
and
\begin{align}
J_1(b)^{CZ} = \frac{ (1 -b^2) (1 + 10b + b^2) 
    + 6b(1+b)^2 \log{b} }{(1-b)^5}.
\end{align}

Lattice results appear to contraindicate the Chernyak-Zhitnitsky form but do suggest that the distribution amplitude is broader than the asymptotic one~\cite{LatticeParton:2022zqc}.

We now defer numerical results until we have discussed alternative SIDIS processes for producing pions.  This is in the next section, and and the point of the discussion is to be able to evaluate if direct pions indeed dominate over a significant kinematic range.  Numerical results will follow in the section after that.


\section{Competing processes}
   \label{sec:other}



\subsection{Vector meson dominance}

\begin{figure}[t!]
\includegraphics[width=0.7\columnwidth]{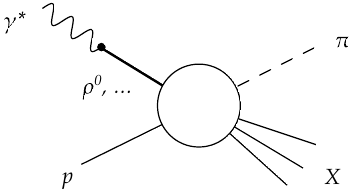}

\caption{Pion production from a vector meson dominated (VMD) virtual photon.} 
\label{fig:vmd}
\end{figure}

Another process for producing SIDIS pions involves vector meson dominance wherein the photon converts to a vector meson which then scatters hadronically from the target, 
Fig.~\ref{fig:vmd}.   We give here only a brief account of how to calculate this contribution to pion production, with details available in~\cite{Afanasev:1999xk}.  This work heavily uses other authors phenomenological fits to meson-proton scattering data, with references given there..

We work first only with the $\rho^0$ meson, and include other vector mesons later.  The $\gamma$-$\rho$ coupling is $e m_\rho^2/f_\rho$, where $f_\rho$ can be obtained from the $\rho^0 \to e^+ e^-$ decay rate~\cite{Pautz:1997eh}.  At the cross section level,
\begin{eqnarray}
 &d\sigma& (\gamma p \rightarrow \pi^+ X) 
  = \frac{\alpha}{\alpha_\rho} 
  \left( \frac{m_\rho^2}{m_\rho^2-q^2} \right)^2
    d\sigma (\rho^0 p \rightarrow \pi^+ X)
                            \nonumber \\
    &+& {\rm other\ VMD} + {\rm non\ VMD\ contributions}  .
\end{eqnarray}

We need the cross section for $\rho^0$ induced $\pi^+$ production.  This is a strong interaction and so a strong interaction sized process.  It can be estimated from various (pion charge changing, to avoid leading particle effects) pion induced inclusive pion production processes, abetted sometimes by porootn induced porcesses to enhance the angular coverage of the data.  Again, details are in~\cite{Afanasev:1999xk}, including treating contibutions from other vector mesons and giving the parameters and functional forms that underlie the VMD results (cf.~also~\cite{Szczurek:2000mj}).


\subsection{Fragmentation}


\begin{figure}[t!]
\includegraphics[width=0.9 \columnwidth]{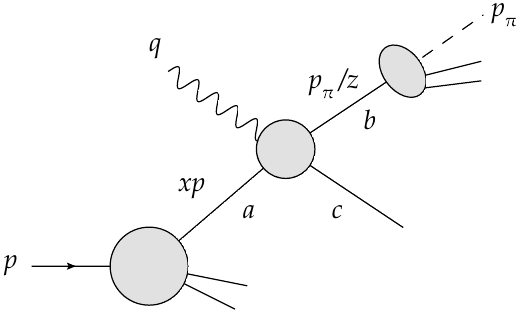}

\caption{Basic diagram for a fragmentation process.} 
\label{fig:fragbasic}
\end{figure}

Since in this paper we explore the fragmentation process only to see where it fades by a significant factor relative to the direct process, we content ourselves with a simple low order treatment.  We need quark distributions, fragmentation functions, and subprocess cross sections.  We use reasonably modern quark distributions (e.g.,~\cite{Gluck:1995yr}) and simple fragmentation functions.  The relevant subprocess cross sections are well known for real photons, and we below give for completeness the versions for virtual photons.

In a fragmentation process, Fig.~\ref{fig:fragbasic}, the virtual photon interacts with parton $a$ in the target and parton $b$ ``fragments'' to produce a pion with some accompanying unobserved debris.  Parton $a$ carries (light-front longitudinal) momentum fraction $x$ of the target hadron, and fraction $z$ of the momentum of the fragmenting parton $b$ goes into the observed pion momentum.  

The cross section for this contribution to the $\gamma^* p \to \pi X$ process is
\begin{align}
\frac{d\sigma}{dx dt dz} &=
\sum_{abc} G_{a/p}(x) \frac{1}{z}
\frac{d\hat\sigma (\gamma^* a \to b c)}{d \hat t}
D_{\pi/b}(z)    .
\end{align}
$G_{a/p}$ is a parton distribution function, or the number of partons of type $a$ (quark or gluon) in interval $dx$, the fragmentation function $D_{\pi/b}$ gives the number of pions produced from parton $b$ per interval $dz$, and $\hat t$ is defined below.  

This can be worked into 
\begin{align}
\omega_\pi \frac{d\sigma}{d^3k} &=
\frac{2 m_p | \vec q \,|}{ -\pi t}
\int_{z_\text{min}}^1 \frac{dz}{z}
\frac{x^2}{1-(1-z)q^2/t}        \nn\\
&\times     \sum_{abc}    G_{a/p}(x) 
\frac{d\hat\sigma}{d \hat t}    D_{\pi/b}(z) ,
\end{align}
with
\begin{align}
x = \frac{ -t + (1-z) q^2 }
{ z(s-m_p^2 -q^2) + u -m_p^2 }
\end{align}
(which gives Eq.~\eqref{eq:xreal} for $z=1$) and
\begin{align}
z_\text{min} = - \frac{t+u-m_p^2-q^2}{s-m_p^2}
= \frac{s - W^2}{s-m_p^2}  .
\end{align}
In the last equation, $W$ is the mass of the hadronic final state particles other than the pion, and shows that $z_\text{min}$ never exceeds unity (for a proton target).


Simple estimates for fragmentation functions giving pions were used in~\cite{Carlson:1993ys} and are
\begin{align}
D_{\pi^+/u}(z) &= \frac{5}{6} (1-z)^2 +
\frac{5}{6} \frac{ (1-z)4 }{z}  ,   \nn\\
D_{\pi^0/u}(z) &= \frac{5}{12} (1-z)^2 +
\frac{5}{6} \frac{ (1-z)4 }{z}  ,   \nn\\
D_{\pi^-/u}(z) &= 
\frac{5}{6} \frac{ (1-z)4 }{z}  ,
\end{align}
and \textit{mutatis mutandi} for the down quark.

Relevant subprocesses are the QCD Compton process $\gamma^* q \to q g$ and the photon gluon fusion process $\gamma^* g \to q + \bar q$.  Cross sections for these for real photons are well known;  for virtual photons, taking the ``T'' cross section (the one with the average of the transverse polarizations) for comparisons, one has for the Compton process
\begin{align}		\label{eq:T}
&\frac{ d\hat\sigma_T}{ d\hat t }
(\gamma^* q \to q g) = 
	\frac{1}{2} \left(  \frac{ d\hat\sigma_{11}}{ d\hat t } + \frac{ d\hat\sigma_{22}}{ d\hat t }  \right)     \nn\\
& \ \  =  \frac{4}{3} 
\frac{2 \pi e_q^2 \alpha \alpha_s }
{ ( \hat s - q^2)^2 }
		\left( - \frac{ \hat t }{ \hat s } - \frac{ \hat s }{ \hat t } - \frac{ 2 \hat u q^2 }{ \hat s \hat t }
			+ \frac{ 2 \hat u q^2 }{ (\hat s - q^2 )^2 }\right)  .
\end{align}
The Mandelstam variables for this subprocess are
\begin{align}
\hat s &= (x p + q)^2  ,  \nn\\
\hat t &= (q - k/z )2  ,  \nn\\
\hat u &= (xp - k/z)^2  .
\end{align}
(Generally, $\hat t$ is the momentum transfer squared between the photon and the parton that will produce the quark, and we have written~\eqref{eq:T} for the case that the quark produces the pion.)  For photon gluon fusion, 
\begin{align}
&\frac{ d\hat\sigma_T}{ d\hat t }
(\gamma^* g \to q \bar q)    \nn\\
&\ \ =  \frac{1}{2} \frac{2 \pi e_q^2 \alpha \alpha_s }{ ( \hat s - q^2)^2 }
		\left(  \frac{ \hat u }{ \hat t } + \frac{ \hat t }{ \hat u } + \frac{ 2 \hat s q^2 }{ \hat t \hat u }
			- \frac{ 4 \hat s q^2 }{ (\hat s - q^2 )^2 }\right)   .
\end{align}

A selection of numerical results are shown in the next section.


\section{Selected numerical results}
    \label{sec:numbers}


\begin{figure}[t!]
\includegraphics[width=0.9\columnwidth]
{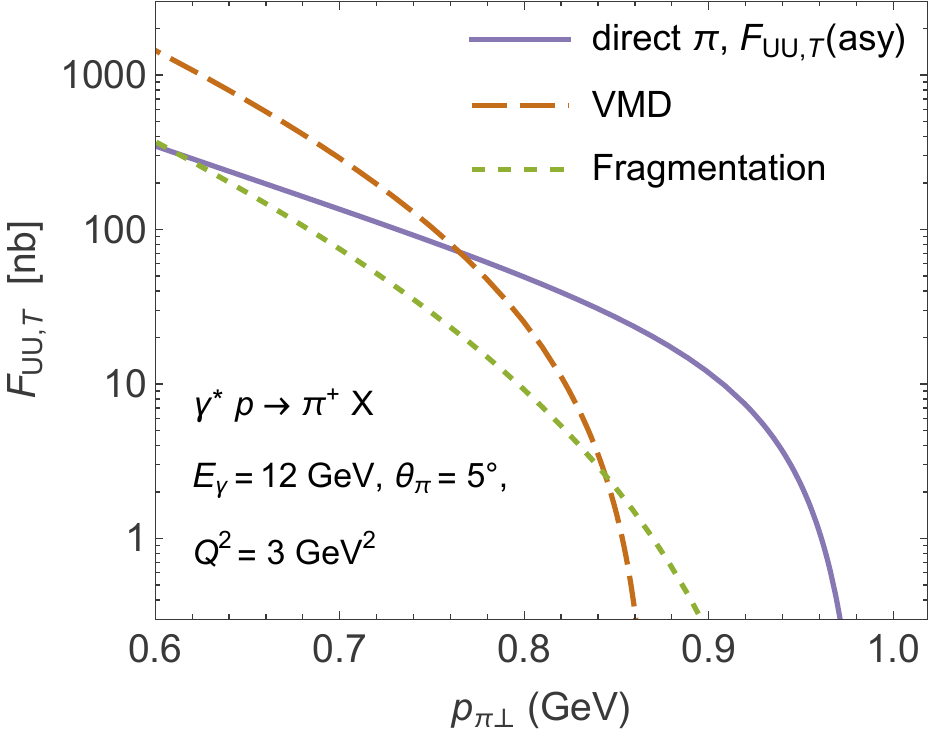}

\bigskip
\includegraphics[width=0.9\columnwidth]
  {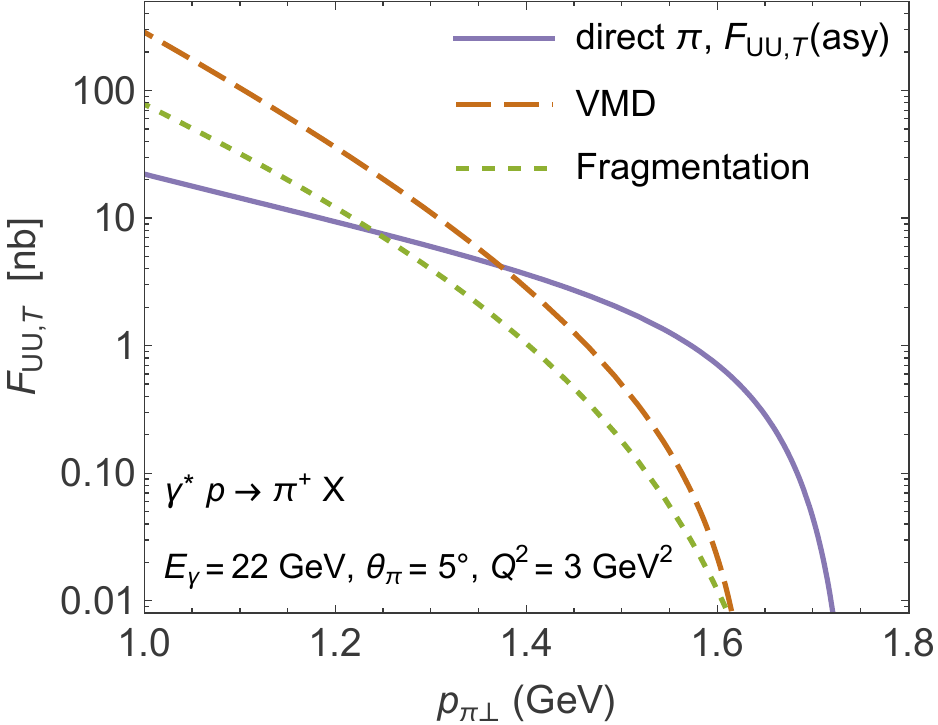}

\caption{A comparison of the rates of pion production from the direct pion, VMD, and fragmentation mechanisms, for two different incoming energies, as a function of pion transverse momentum.  This plot uses the asymptotic distribution amplitude for isolated pion calculation; other choices considered in the text would give more copious isolated pion production.}
\label{fig:comp3}
\end{figure}

The first question is whether the claim that the isolated pion process is dominant for high transverse momentum pion production is true, and if so, at what momenta.  As an answer, we present two plots in Fig.~\ref{fig:comp3}, which show the $\sigma_T$ cross section (which will prove to be the largest of the five cross sections in Eq.~\eqref{eq:xsctn}) for the isolated pion process, the VMD process, and the fragmentation process. The plots are for $E_\gamma = 12$ GeV and $22$ GeV, both with pion lab angle $\theta_\pi=5^\circ$, and $Q^2 = 3$ GeV$^2$.  One sees the isolated process rising above the others for high enough $k_T$, about 0.8 and 1.4 GeV in these two cases.  This plot uses the asymptotic distribution amplitude for the isolated pion calculation.  It gives the smallest results among the distributions amplitudes selected in this study;  the results from the CZ amplitude would be nearly three times larger and would show dominance of the isolated pion mechanism at somewhat lower $p_{\pi\perp}$.   The choice of $Q^2$ matters, as with rising $Q^2$ the isolated pion process cross section rises relative to the other processes.

\begin{figure}[t!]
\includegraphics[width=\columnwidth]{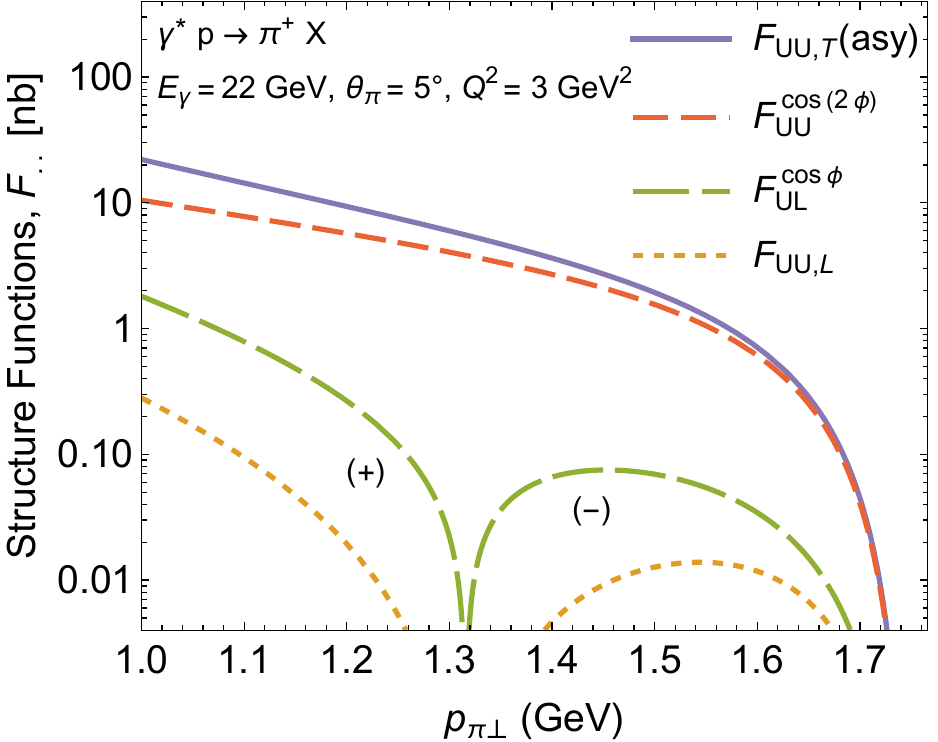}

\caption{Four contributions, with differing input photon polarizations, to isolated pion production. For this case,  $\sigma_{TT}$ is negative, and  both $\sigma_{LT}$ and $\sigma_L$ have a zero at these kinematics.} 
\label{fig:isolated}
\end{figure}

A further question is what are the relative sizes of the $\sigma_T$, $\sigma_{TT}$, $\sigma_L$, and $\sigma_{LT}$ cross sections, and these are plotted in Fig.~\ref{fig:isolated}.
For these kinematics, involving the longitudinal polarization clearly gives a smaller results than the terms involving only the transverse polarizations.  We may remark that the quantity $C$ can sometimes have a zero in the region of interest, as it does here.

\begin{figure}[t!]
\includegraphics[width=\columnwidth]{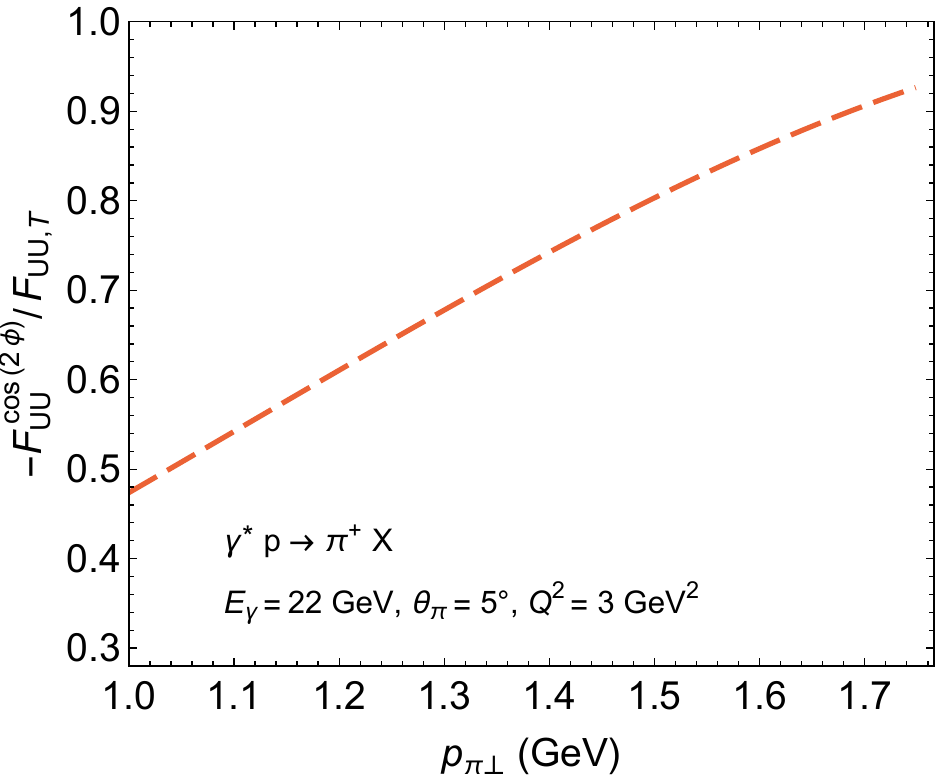}

\caption{A comparison of the $\sigma_{TT}$ and $\sigma_T$ terms, indicating the relatively easy possibility of seeing the $\cos^2 \phi_h$ modulation.}
\label{fig:TTT}
\end{figure}

The TT term, or the difference between the two transverse polarization cross sections, is of course smaller in magnitude than the the T term, which comes from the sum, but still is of comparable size, as shown in Fig.~\ref{fig:TTT}, which plots the absolute value of the ratio $\sigma_{TT}/\sigma_T$.  This speaks to the possible ease of observing the $\cos(2\phi)$ modulation of the overall cross section.  

\begin{figure}[t!]
\includegraphics[width=\columnwidth]{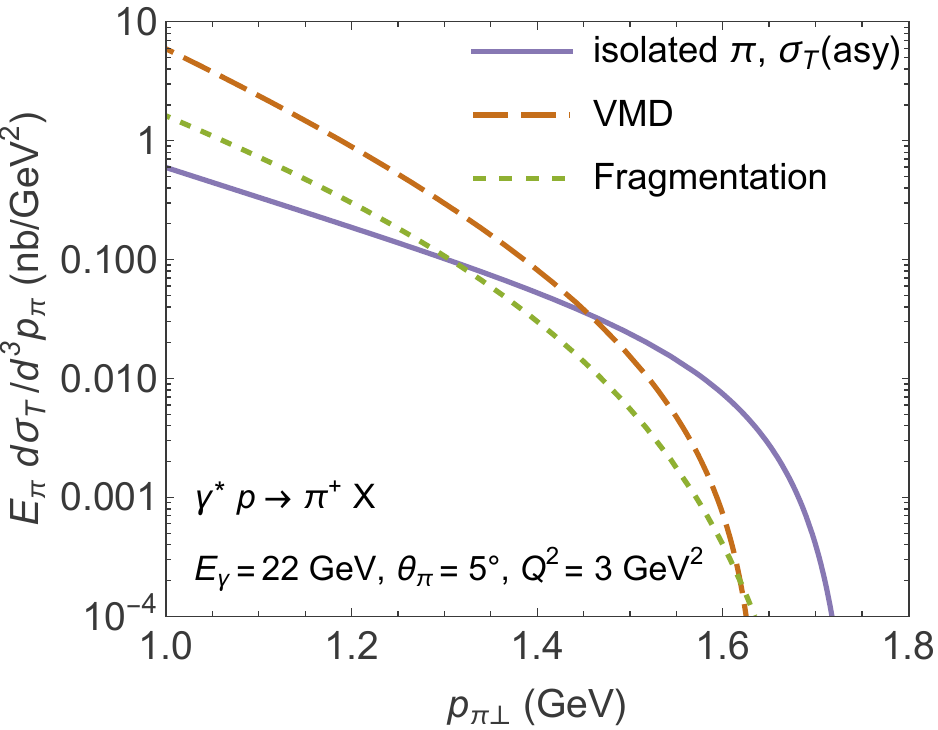}

\caption{This time for the $\pi^0$,  four contributions with differing input photon polarizations to isolated pion production. For $\pi^0$ production, $\sigma_L$ becomes quite prominent.} 
\label{fig:pizeroisolated}
\end{figure}

A corresponding plot showing the sizes of the $\sigma_T$, $\sigma_{TT}$, $\sigma_L$, and $\sigma_{LT}$ cross sections for the neutral pion production is shown in Fig.~\ref{fig:pizeroisolated}.  for the $\pi^0$, the longitudinal cross section is prominant at high $p_{\pi\perp}$, a feature already noted in~\cite{Afanasev:2003ne}.

\begin{figure}[t!]
\includegraphics[width=\columnwidth]{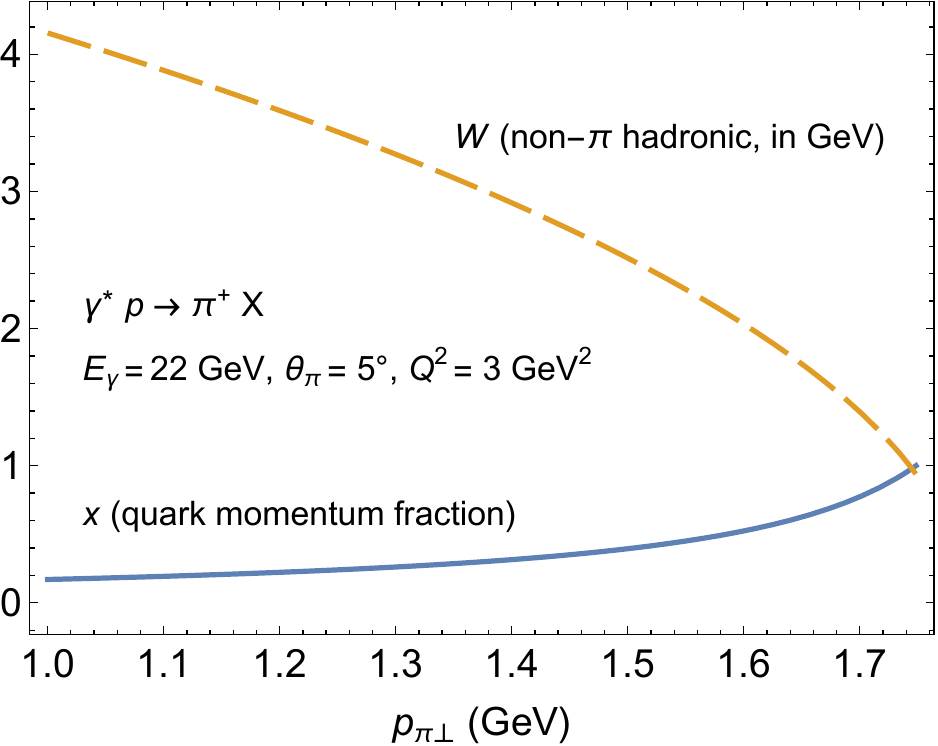}

\caption{The upper curve here shows the invariant mass in GeV of the unobserved hadronic system accompanying the pion.  For $p_{\pi\perp}\lesssim 1.6$ GeV, this mass is above the resonance region (taking the boundary as $2$ GeV).  Recall that for these kinematics, if the CZ distribution amplitude gives the correct normalization,  isolated pion production exceeds VMD production for $p_{\pi\perp} \gtrsim 1.3$ GeV.  The lower curve shows the quark momentum fraction $x$, which ranges from 0.25 to 0.52 as $p_{\pi\perp}$ ranges from 1.3 to 1.6 GeV. } 
\label{fig:wx}
\end{figure}

Finally for this section, we investigate the invariant mass $W$ of the non-observed hadronic material accompanying the pion and also the size of the initial quark momentum fraction $x$ for the isolated pion process.  These are put on the same plot, Fig.~\ref{fig:wx}.  For the $E_\gamma = 22$ GeV, $\theta_\pi=5^\circ$, $Q^2 = 3$ GeV$^2$ kinematics chosen for our examples, $W$ is above the  resonance region, taken to means $W > 2$ GeV, for $p_{\pi\perp} < 1,6$ GeV, which means there is a significant region of isolated pion production where the recoiling hadrons are above the resonance region, avoiding potential final state interactions that could modify some of the results.  Quantity $x$ is on the same plot, and $x$ is above 0.26 and 0.31 for $p_{\pi\perp} =$ 1.3 and 1.4 GeV, respectively.  Hence any relevant measurements would be mainly in the valence quark region.  

\section{Comments on Extracting pdf and DA}
   \label{sec:learn}


Upon establishing a region where direct pion production dominates, one may use the process as a tool to clarify aspects of proton and pion structure.  The direct pion production cross sections are generically given as a quark pdf in the proton (or other chosen target) times a $q^2$ and $t$ dependent integral involving the pion's (or in the future, other meson's) distribution amplitude time known and well determined kinematic factors, roughly
\begin{align}
    d\sigma = f_q(x) \cdot J_1(q^2/t) \cdot
        \text{ kinematic factors} .
\end{align}

By making suitable kinematic choices, one can isolate the $x$ dependence of the pdf or the differences in $J_1$ due to different choices of the DA.  For the virtual photon subprocess, there are 4 kinematic variables, two for the the photon($\nu = E_\gamma$ and $q^2$) and two for the pion (for example, $E_\pi$ and $\theta_\pi$ in the lab frame).  One can fix three linear combinations of the variables stated and vary a fourth to see sensitivities to the pdf and DA.  As a side remark, the White Paper regarding 22 GeV electrons at JLab~\cite{Accardi:2023chb} 
emphasizes the ability to obtain extensive multidimensional data, so free choices in data analysis should abound.  We shall note a few examples.  

\begin{figure}[t!]
\includegraphics[width=\columnwidth]{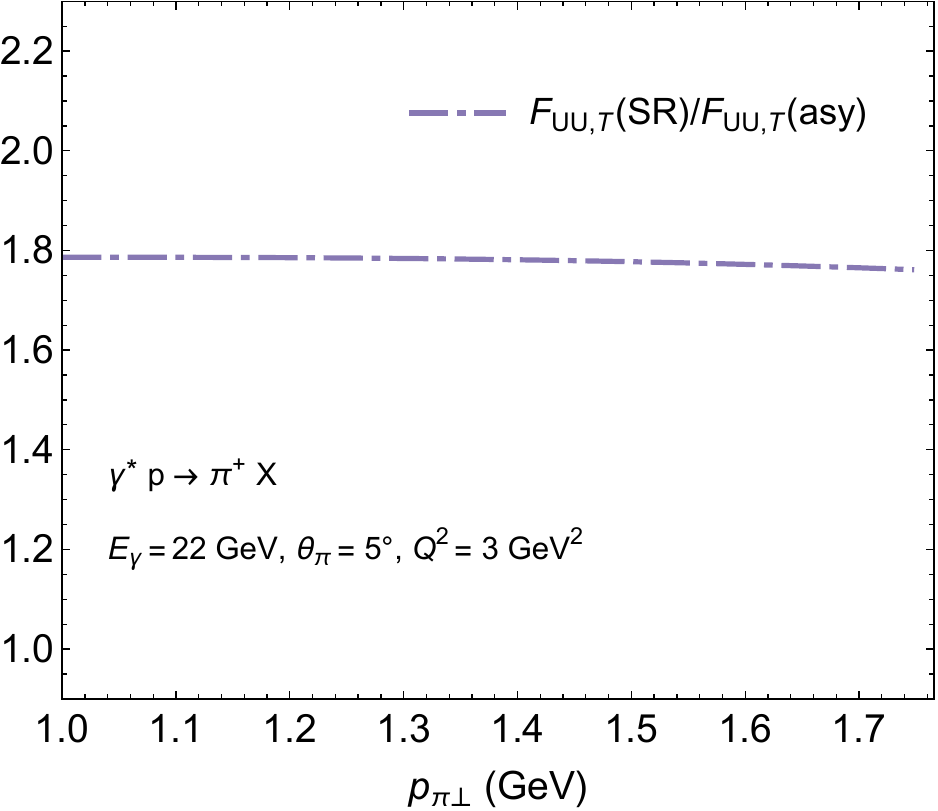}

\caption{Ratio of the transverse cross sections for pion electroproduction, at the selected kinematics, for the square root (SR) and asymptotic DAs.} 
\label{fig:ratioTpiplus}
\end{figure}

For studying the $u$ quark valence pdf, a plot already shown, Fig.~\ref{fig:isolated}, provides a good beginning.  This plot has $E_\gamma$, $Q^2$, and $\theta_\pi$ fixed.  Plot of Fig.~\ref{fig:ratioTpiplus} shows how the integral changes when switching between pion DA and another; more explicitly it shows the ratio of $d\sigma_T/d^3p_\pi$ calculated for two different DAs and potted for the same kinematics as Fig.~\ref{fig:isolated}.   The normalization changes, but the curve is nearly flat.  That means that the shape of the experimental curve for $\sigma_T$ is determined by the shape of the $q_u(x)$ pdf.  The curve actually shown in Fig.~\ref{fig:isolated} was calculated using the GRSV parton distribution.  Data will pick out whatever parton distribution is correct.  

\begin{figure}[t!]
\includegraphics[width=0.9\columnwidth]{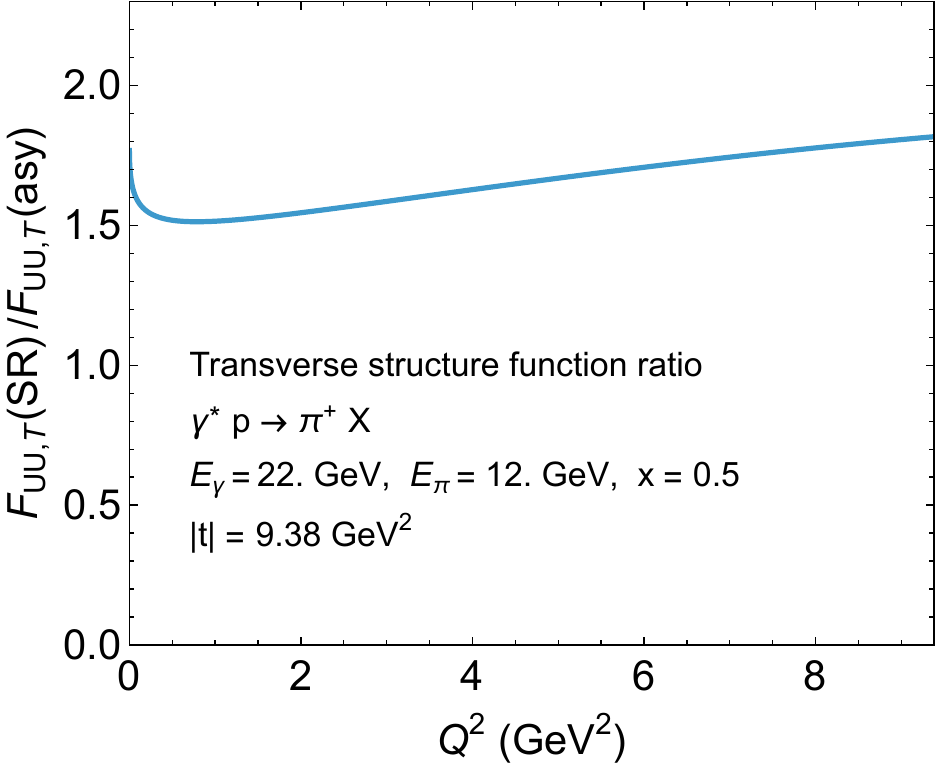}

\medskip

\includegraphics[width=0.9\columnwidth]{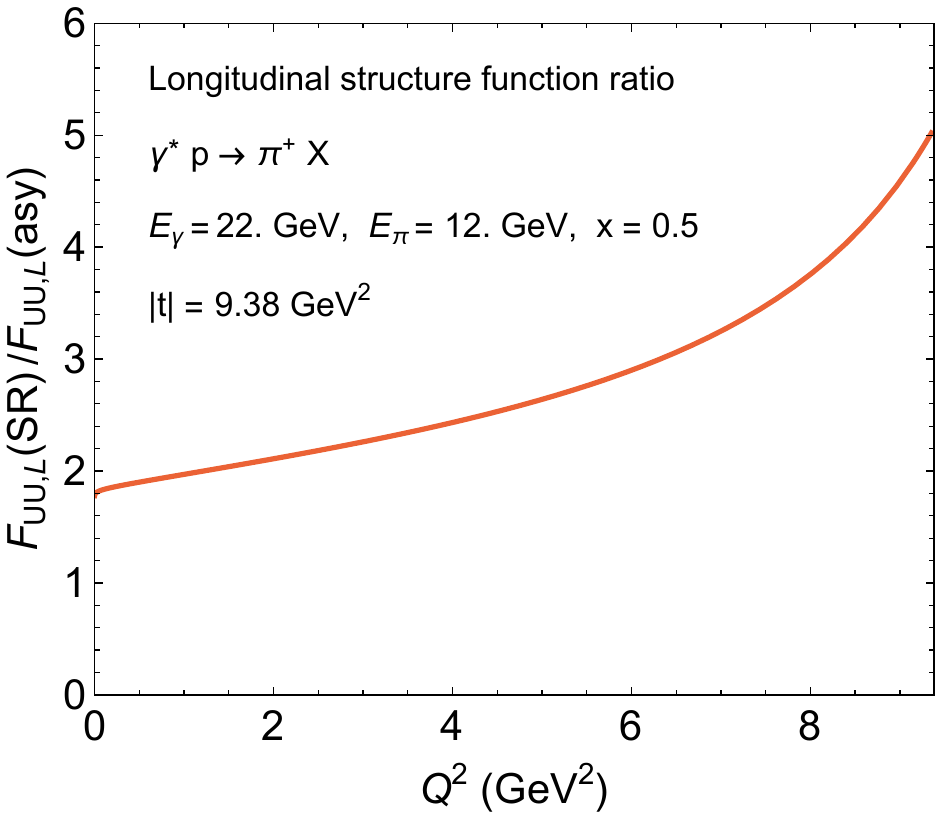}

\caption{Ratios of the transverse cross sections (upper panel) and longitudinal cross sections (lower panel) for $\pi^+$ electroproduction, at these  kinematics, again for the square root (SR) and asymptotic distribution amplitudes.} 
\label{fig:ratioTLpiplus}
\end{figure}

To learn more about the pion DA, one can search for other suitable kinematics, for example fixing $x$ and also fixing $E_\gamma$ and $E_\pi$, which one can check also fixes $t$.  Choosing $Q^2$ as the remaining variable allows significant variation in the argument of $J_1(q^2/t)$ and leads to the ratio plots for different DAs shown in Figs.~\ref{fig:ratioTLpiplus}.  The result for the transverse cross section is relatively flat and hence not so interesting, but the result for the longitudinal case show a significant variation as $q^2$ goes from 0 to approximately $t$.  Hence there is an opportunity to distinguish different DAs.  With enough data an actual inversion of $J_1(q^2/t)$ to obtain $\phi_\pi(y)$ could be assayed.

\begin{figure}[t!]
\includegraphics[width=0.9\columnwidth]{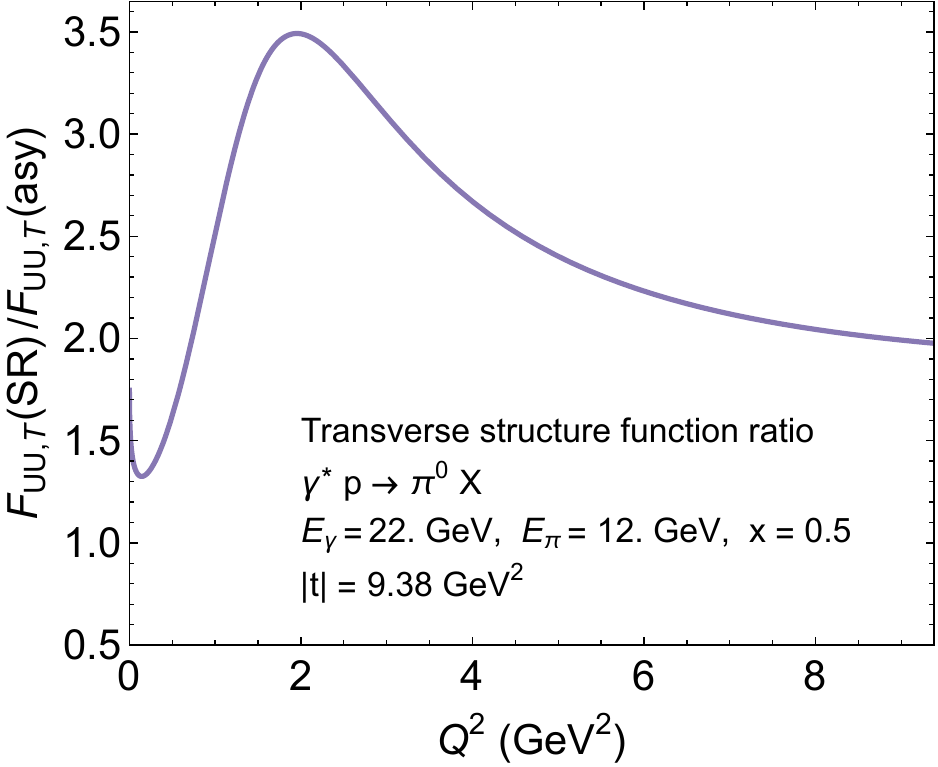}

\medskip

\includegraphics[width=0.9\columnwidth]{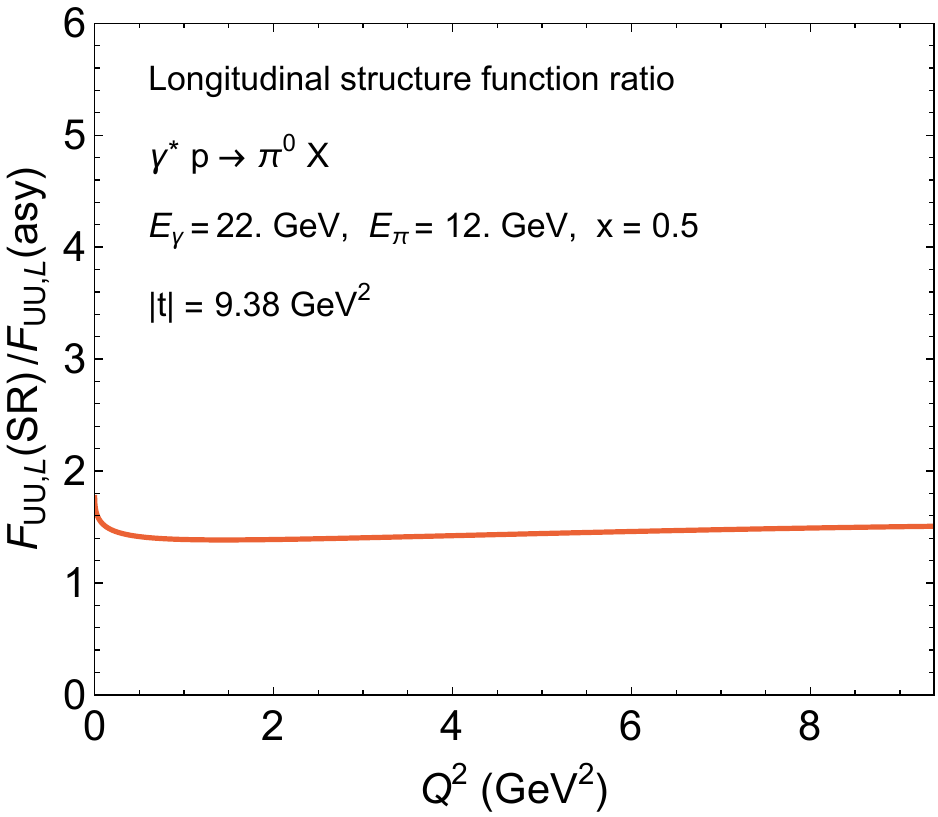}

\caption{Ratios of the transverse cross sections (upper panel) and longitudinal cross sections (lower panel) for $\pi^0$ electroproduction, at these  kinematics, again for the square root (SR) and DAs.} 
\label{fig:ratioTLpizero}
\end{figure}

Some corresponding plots for the $\pi^0$ case are in Fig.~\ref{fig:ratioTLpizero}.  Here the transverse cross section appears to be the more shape sensitive.


\section{Final remarks}
   \label{sec:end}


We have calculated direct, or isolated, isolated pion electroproduction from proton targets, by methods easily extended to other hadronic targets.  We have showed that in electroproduction at moderate energies the isolated pion process is dominant over a significant region of high momentum pions.

We have also noted that the calculation is proportional to the parton distribution functions at high momentum fraction and the meson's distribution amplitude, with other factors entering the calculation known and established.  Hence the measured cross sections can allow us to learn the pdfs and DAs more precisely.  We have shown by example that there are kinematic choices that will allow learning the shape of the quark pdf whatever the meson's DA, and other choices that will gather information about the meson's DA whatever the quark distribution function.  These examples also show the value of data over a wide kinematic range, including data with transverse and longitudinal contributions separated. 

The perturbative QCD mechanism for direct pion production in SIDIS considered here is closely related to a short-range part of the scattering amplitude for exclusive electroproduction of pions at high transferred momenta, and is of interest to the studies of Generalized Parton Distributions of a nucleon \cite{DIEHL200341, ahmad2009nucleon, goloskokov2010attempt, goloskokov2011transversity}. 

There are further investigations that can follow upon the present work that we will plan to investigate.  These include, but may not be limited to, the effect of radiative corrections, particularly including electron bremsstrahlung that affects the relation of the meson's transverse momentum to the internal photon's direction~\cite{[][ and further work to come.]Liu:2019srj};  the inclusion of target proton (or other target) polarization, so that the be 18 structure functions instead of five;  and performing the corresponding calculation for direct rho meson production, which in turn leads to high transverse momentum dipion production.


\section*{Acknowledgements}

A.A.~thanks the National Science Foundation (USA) for support under grant PHY-2111063 and PHY-2514669. C.E.C.~thanks the National Science Foundation (USA) for support under grant PHY-1812326.  We thank Joe Karpie, Jianwei Qiu and Harut Avakian for useful conversations.


\newpage 

\bibliography{isopionrefs}

\end{document}